\documentclass[twocolumn]{aastex631}

\begin{document}

\title{Satellite Constellation Avoidance with the Rubin Observatory Legacy Survey of Space and Time}

\author[0000-0002-8400-1910]{Jinghan Alina Hu}
\affiliation{Harvey Mudd College, Claremont, CA, USA}
\author[0000-0003-1305-7308]{Meredith L. Rawls}
\affiliation{Department of Astronomy / DiRAC / Vera C. Rubin Observatory, University of Washington, Seattle, WA, USA}
\author[0000-0003-2874-6464]{Peter Yoachim}
\affiliation{Department of Astronomy / DiRAC / Vera C. Rubin Observatory, University of Washington, Seattle, WA, USA}
\author[0000-0001-5250-2633]{\v{Z}eljko Ivezi\'{c}}
\affiliation{Department of Astronomy / DiRAC / Vera C. Rubin Observatory, University of Washington, Seattle, WA, USA}

\begin{abstract}
We investigate a novel satellite avoidance strategy to mitigate the impact of large commercial satellite constellations in low-Earth orbit on the Vera C. Rubin Observatory Legacy Survey of Space and Time (LSST). We simulate the orbits of currently planned Starlink and OneWeb constellations ($\sim$40,000 satellites) to test how effectively an upgraded Rubin scheduler algorithm can avoid them, and assess how the overall survey is affected.
Given a reasonably accurate satellite orbit forecast, we find it is possible to adjust the scheduler algorithm to effectively avoid some satellites. Overall, sacrificing 10\% of LSST observing time to avoid satellites reduces the fraction of LSST visits with streaks by a factor of two. Whether such a mitigation will be required depends on the overall impact of streaks on science, which is not yet well quantified. This is due to a lack of adequate information about satellite brightness distributions as well as the impact of glints and low surface brightness residuals on alert purity and systematic errors in cosmological parameter estimation. A significant increase in the number of satellites or their brightness during Rubin Operations may make implementing this satellite avoidance strategy worthwhile.
\end{abstract}



\section{Introduction} \label{sec:intro}

Rubin Observatory's Legacy Survey of Space and Time (LSST) is a ten-year astronomical imaging survey that will begin in 2024 from a new telescope under construction in Chile. Instead of soliciting individual requests for what the telescope should observe, the LSST will uniformly survey the sky every few nights using six color filters to create a decade-long high-resolution survey of the entire southern sky, and share massive quantities of data products with the astronomy community \citep{overview}. To accomplish this, the LSST will employ a scheduling algorithm that uses a modified Markov Decision Process which can generate lists of desirable observations in real time \citep{naghib19}. The LSST scheduler balances the desire to minimize slew time, optimize signal to noise in individual images, and to maintain survey footprint uniformity.

One challenge for the LSST is that increasing numbers of bright low-Earth orbit (LEO) satellites (e.g., Starlink) are being launched, which may leave streaks in astronomical pointings. LEO satellites are visible from Earth because they reflect sunlight, especially during twilight. As the Sun-illuminated satellites move across the field of view of an astronomical pointing, they leave a streak in the image. While the flux from satellite streaks can in many cases be identified and removed, the resulting pixels have much lower signal-to-noise. For a thorough discussion of the scientific utility of residual light after masking satellite trails, see \citet{hasan22}. Over the last three years, many astronomers have raised concerns about the impact of the proliferation of commercial satellites on the LEO ecosystem and astronomical surveys \citep{lawrence22,tyson20}. In addition, astronomers have come together with satellite operators and other stakeholders to create recommendations and strategies to mitigate impacts to observational astronomy and beyond \citep{satcon1,satcon2,dqs1,dqs2}.

\citet{tyson20} used a very simple algorithm to see if the LSST could avoid imaging satellite streaks. They concluded that attempting to naively dodge of order 48,000 LEO satellites is useless because it is operationally inefficient.
In this paper, rather than try to avoid all satellite streaks, we incorporate satellite avoidance as a component of the LSST scheduler's Markov Decision Process. This allows us to avoid a significant fraction of satellite streaks and investigate what level of avoidance might be acceptable because it does not drastically impact the overall performance of the LSST.

There are other efforts underway to mitigate the impact of satellite streaks in astronomical images. For example, satellite companies like SpaceX have worked on darkening the exterior of satellites so they will be less visible\footnote{\url{https://api.starlink.com/public-files/BrightnessMitigationBestPracticesSatelliteOperators.pdf}}. However, even with the most effective darkening mitigations to date, satellites still appear bright to the LSST Camera, and are likely to cause effects like non-linear crosstalk or glints that are challenging to correct with the LSST Science Pipelines software and may introduce systematic biases or spurious detections. This is discussed in more detail in \citet{tyson20} and on the Rubin Observatory LSST Project website\footnote{\url{https://ls.st/satcon}}. Astronomers have also developed algorithms for masking satellite trails in images, but covering the outer wings of the trails without losing extra pixels remains a challenge \citep{hasan22}. The rapid increase in population of LEO satellites threatens to compromise the quality and scientific value of LSST images and also requires extra human and computer resources to effectively mask trails. Thus, we explore an additional option: incorporating the orbits of known commercial satellites into the LSST scheduler so the worst of them may be avoided.

In this paper, we first create realistic simulated forecasts of satellite orbits in Section \ref{method}. We then build a tool that uses that data to create new scheduler constraints, and test the impact of the modified scheduler algorithm on LSST observing programs in Section \ref{results}. Finally, we discuss the resulting trade-off between number of streaks and reduced survey depth that results from avoiding satellites in Section \ref{discuss}, and lay out possibilities for the future as the satellite population changes during Rubin Operations. We also make available a GitHub repository with the data and software necessary to reproduce the paper's figures\footnote{\url{https://github.com/lsst-sims/satellite-dodging-ApJL}}.

\section{Methods}\label{method}

\begin{figure*}[ht!]
\epsscale{0.35}
\plotone{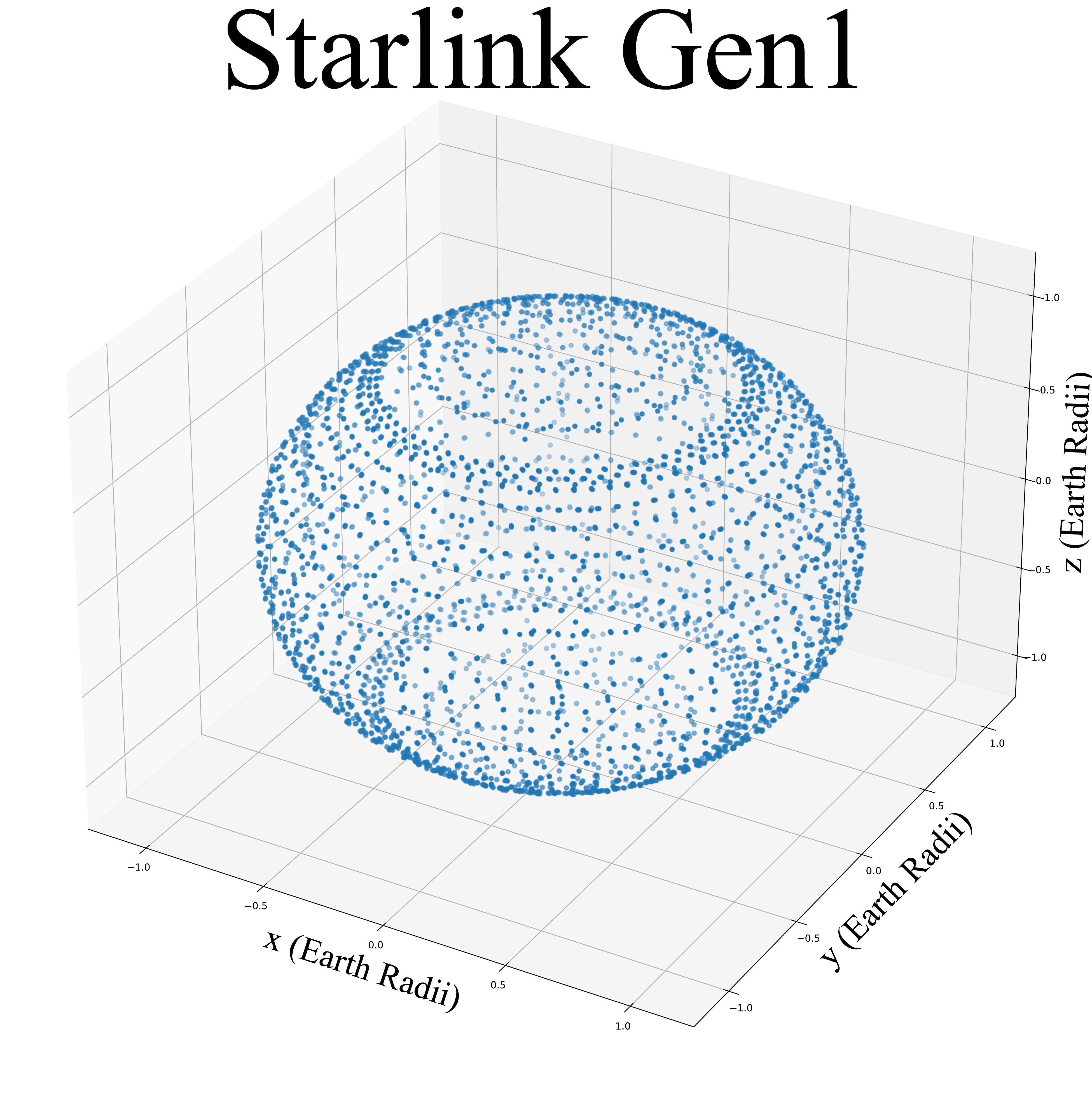}
\plotone{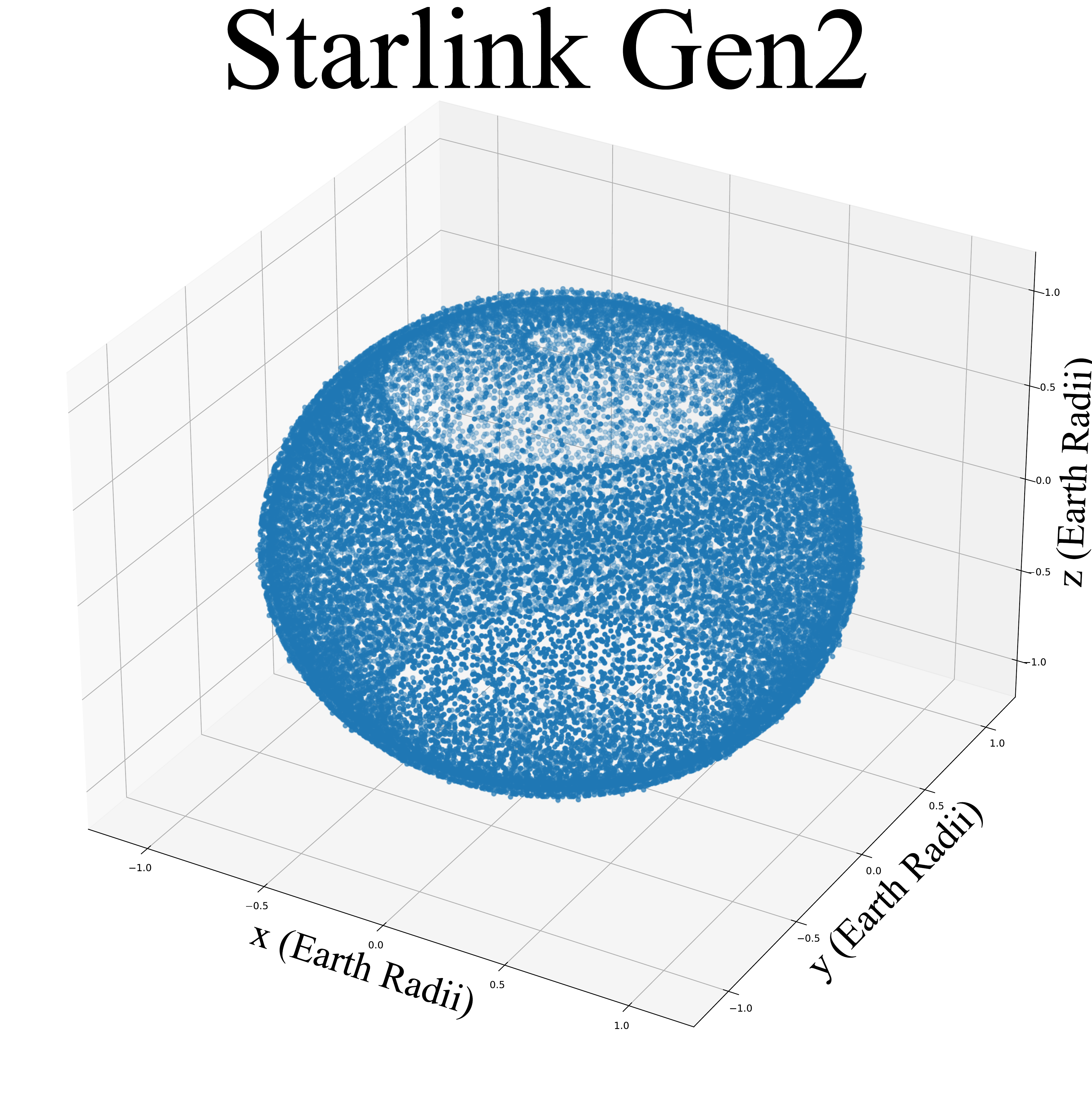}
\plotone{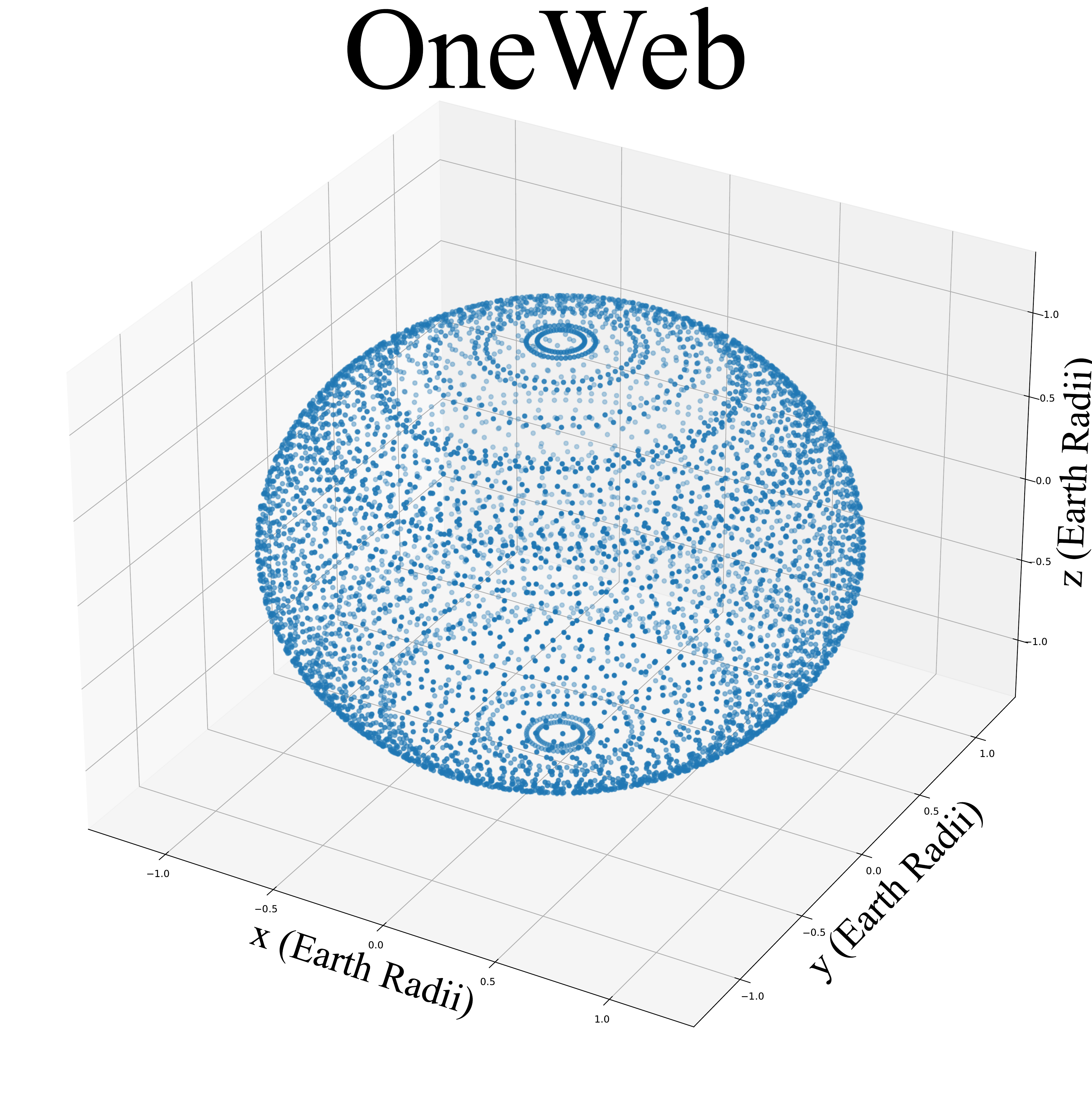}\\
\plotone{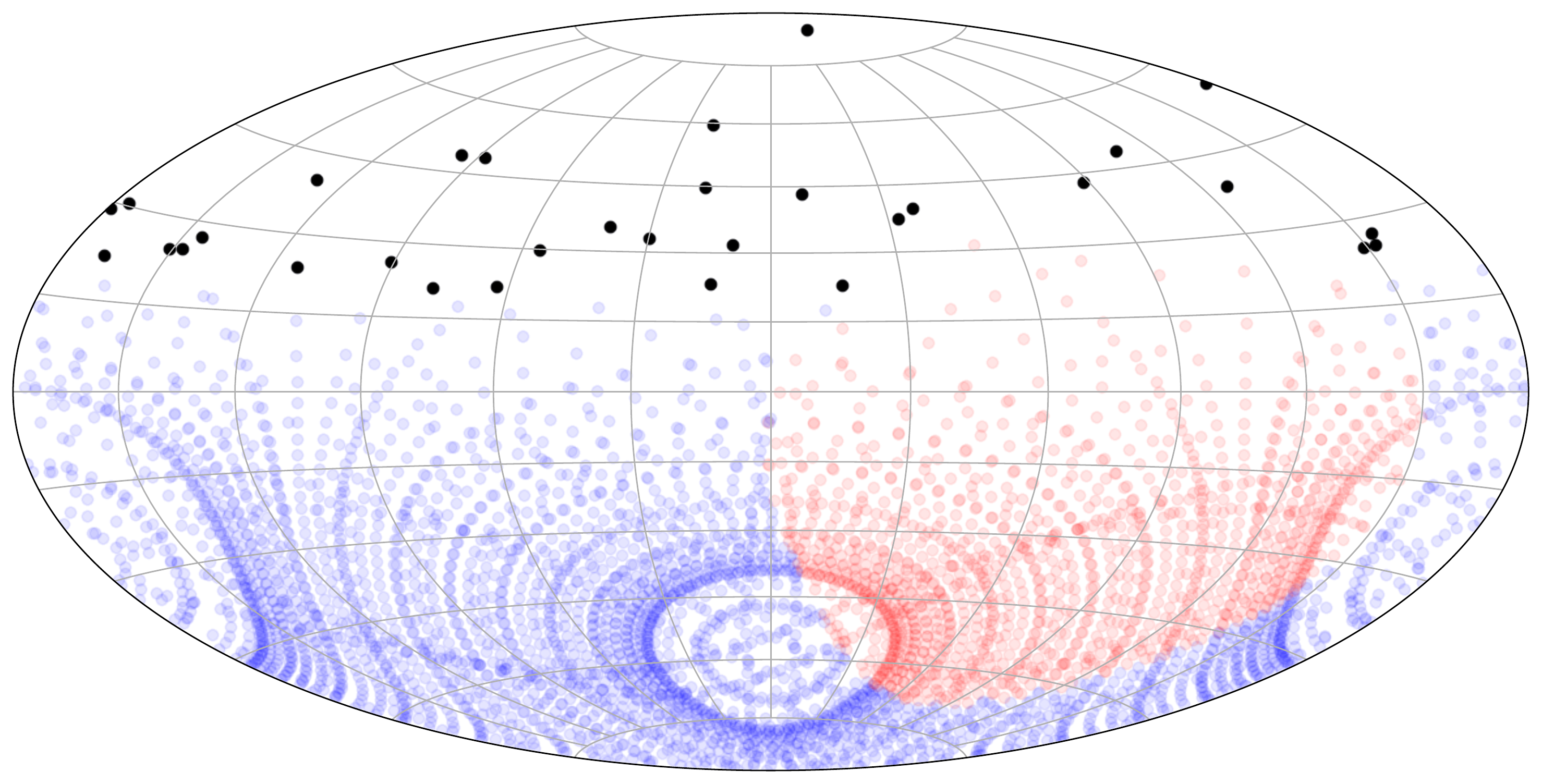}
\plotone{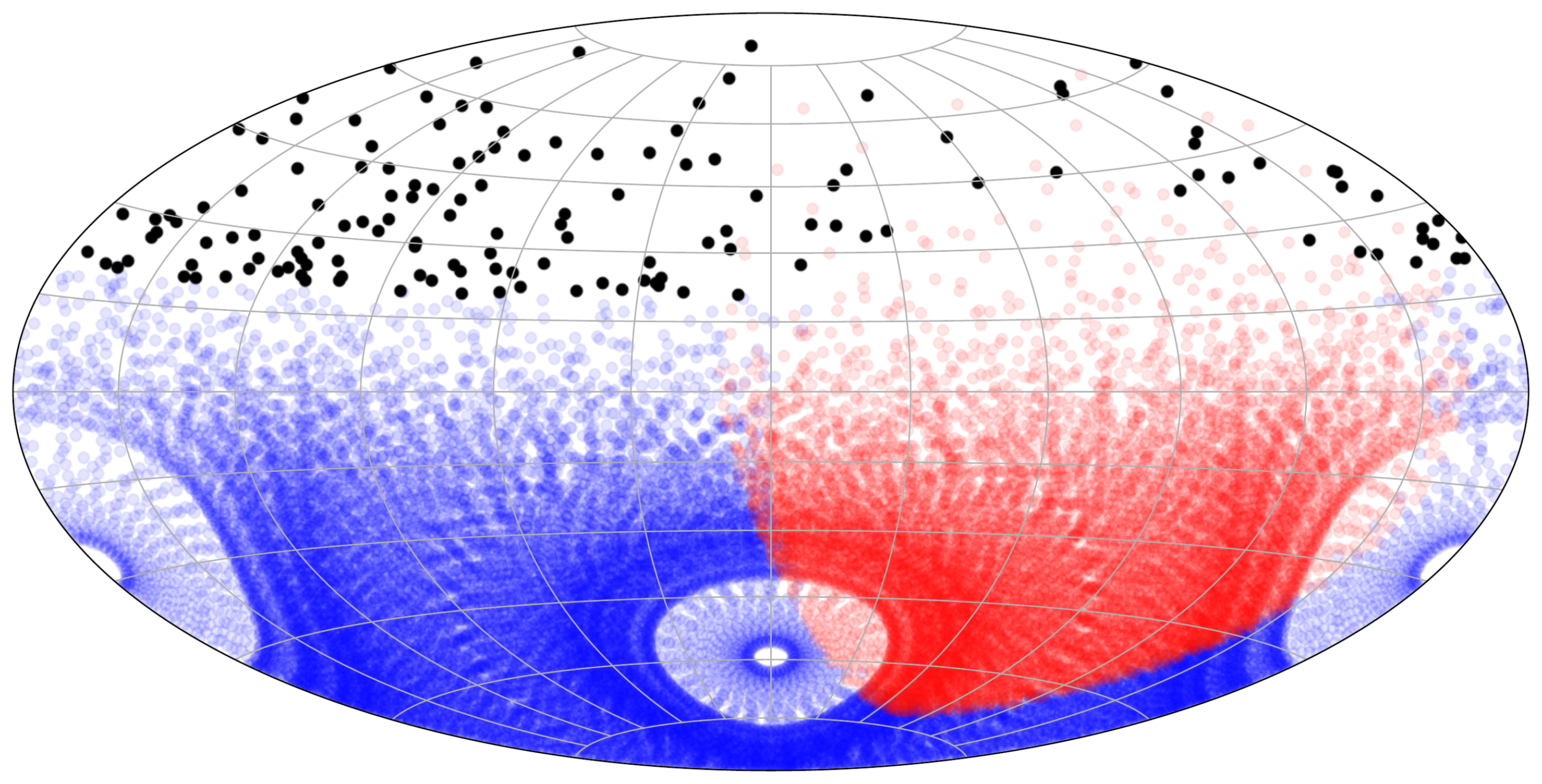}
\plotone{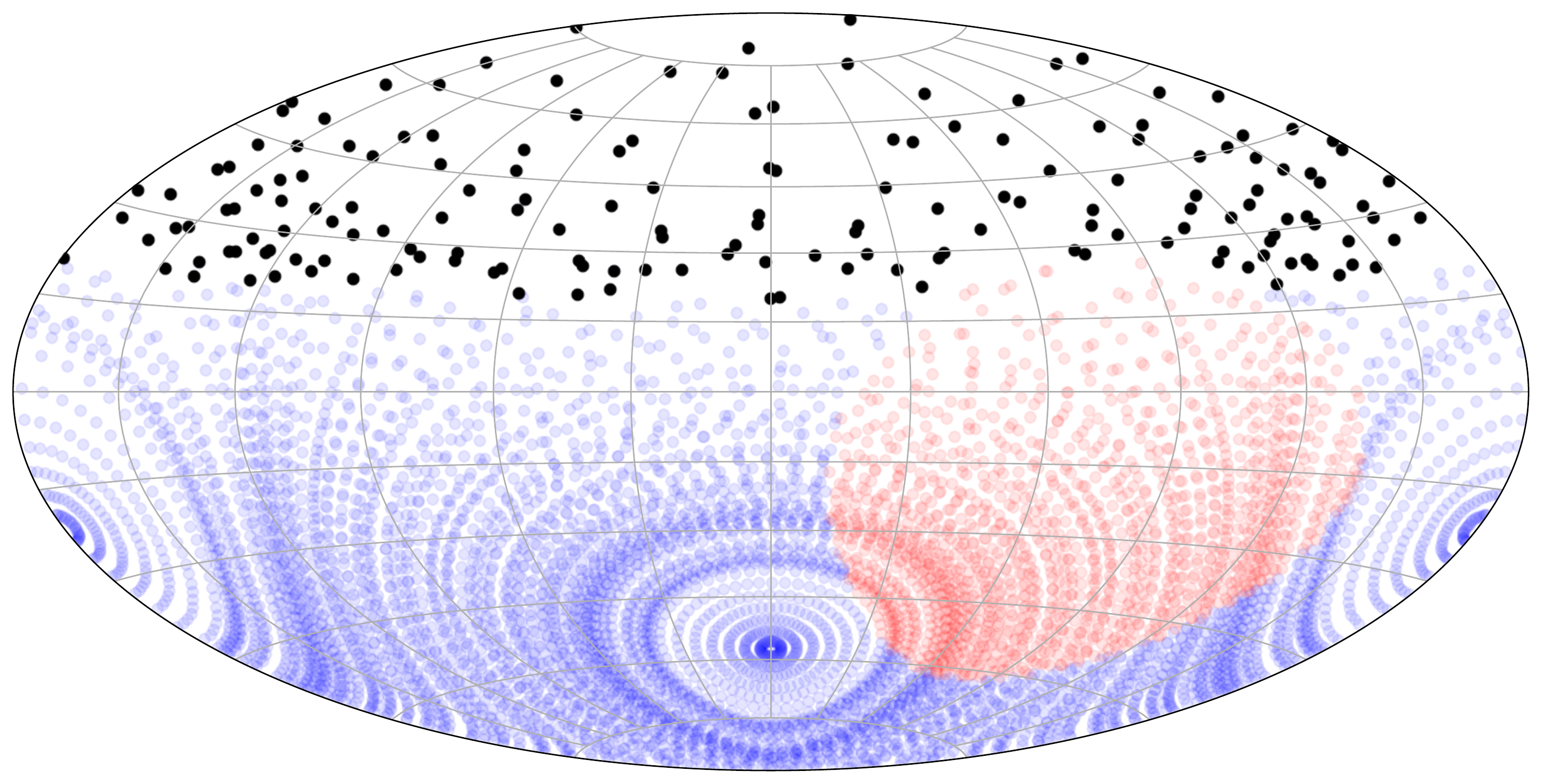}\\
\plotone{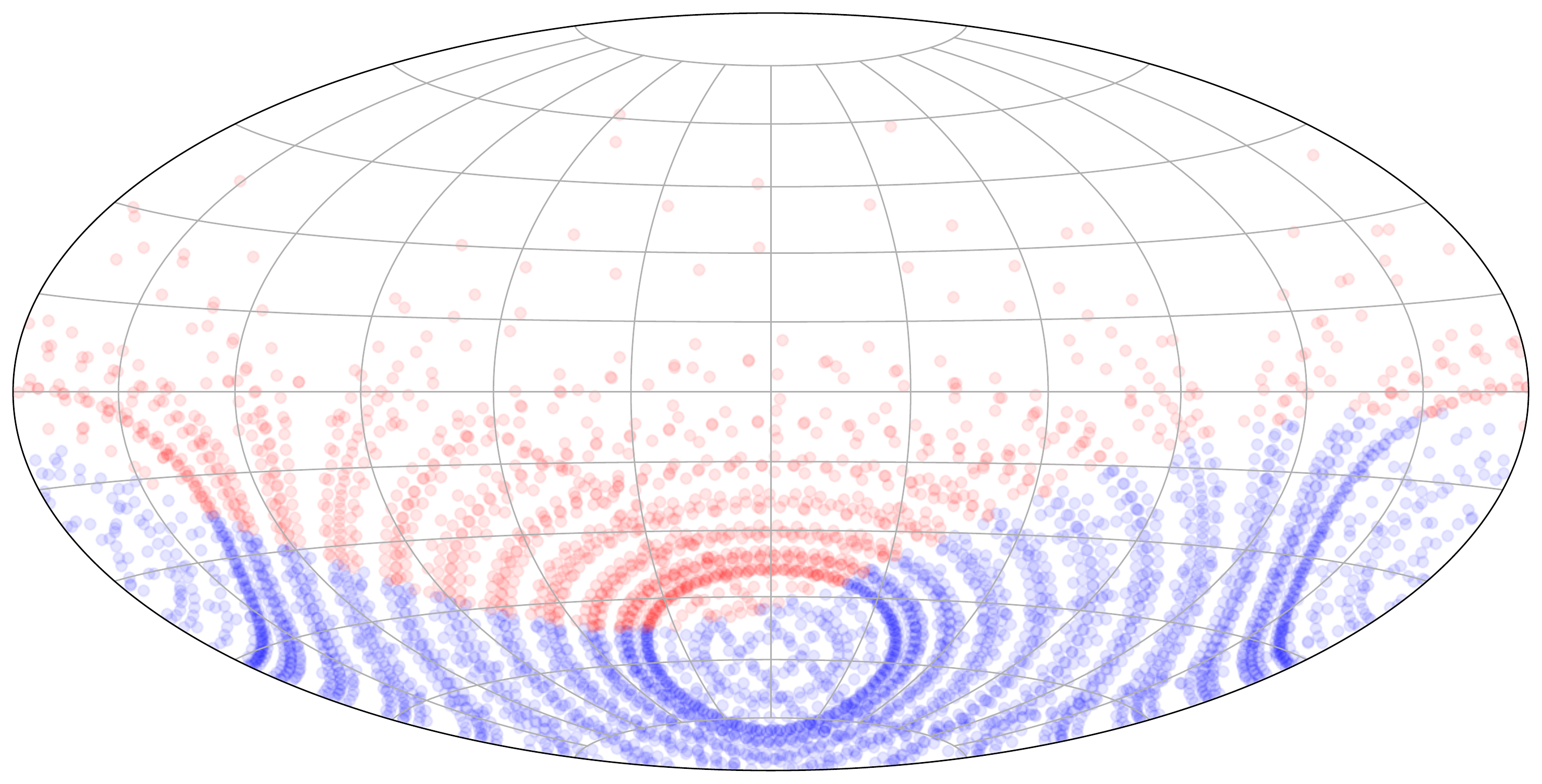}
\plotone{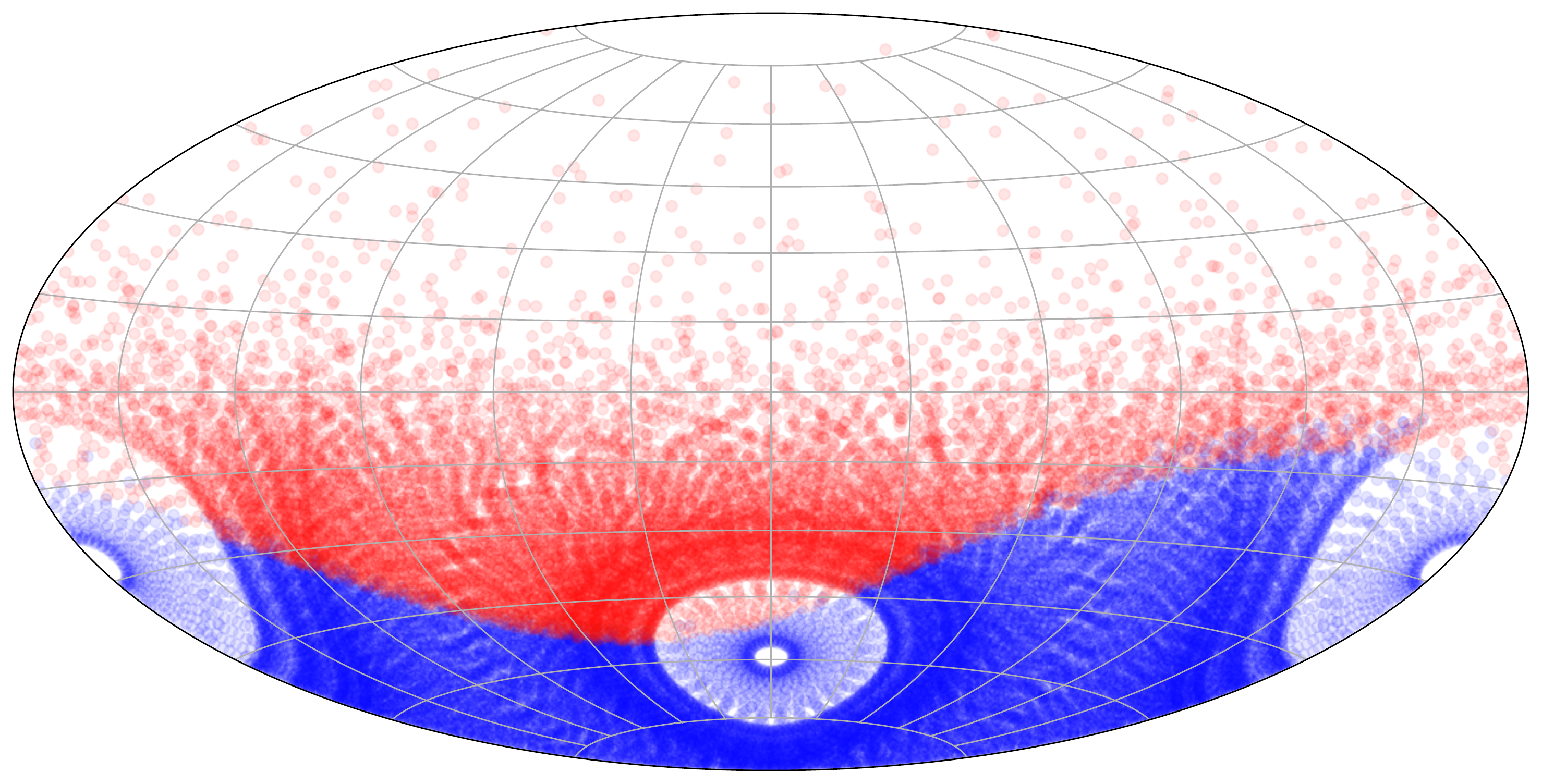}
\plotone{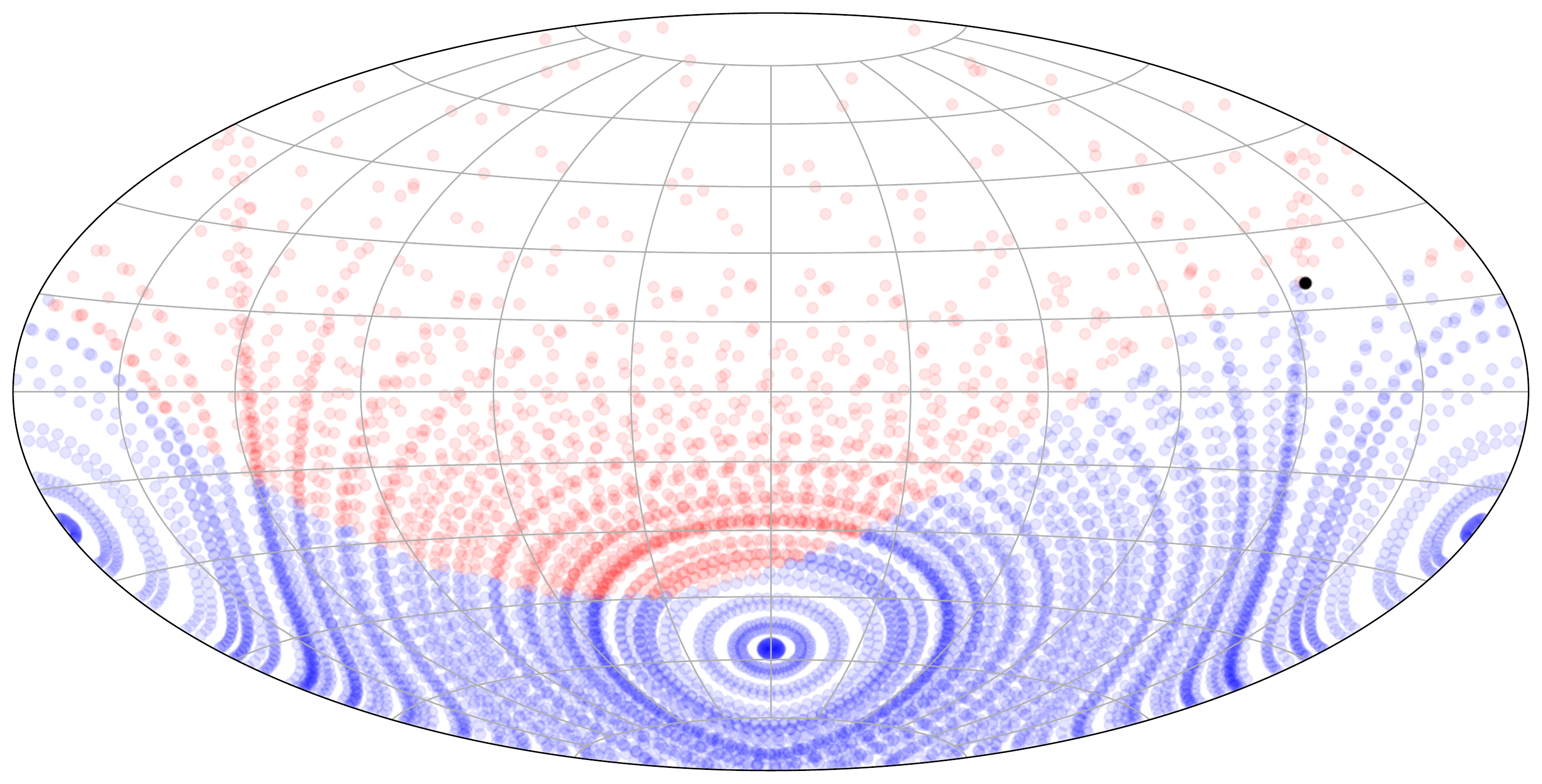}
\epsscale{1}
\caption{Three simulated satellite constellations, one per column. Starlink Gen1 is 4,408 satellites, Starlink Gen2 is 29,988 satellites, and OneWeb is 6,372 satellites, for a grand total of 40,768. The top row shows the 3D distribution of each constellation around Earth. The middle row shows an instantaneous Hammer projection of the altitude and azimuth positions of each constellation as seen from Rubin Observatory on October 1, 2023 during twilight (Sun altitude $-18$ degrees). Blue points are satellites illuminated by the Sun at this time, red points are satellites not illuminated by the Sun, and black points are satellites that are both illuminated and above the Rubin 20 degree altitude pointing limit. The bottom row is the same Hammer projections six hours later in the middle of the night (Sun altitude $-50$ degrees). Because Starlink satellites orbit at 550 km, none are illuminated in the middle of the night at this time of year. The OneWeb constellation at 1200 km has only a single illuminated satellite above the Rubin altitude limit at this particular time.
\vspace{4em}
\label{fig-simulated-constellations}
}
\end{figure*}

We begin by creating realistic forecasts of three commercial satellite constellations, which are illustrated in Figure~\ref{fig-simulated-constellations}. These are Starlink Gen1 (4,408 satellites, altitude $540-570$ km), OneWeb (6,372 satellites, altitude 1200 km), and Starlink Gen2 (29,988 satellites, altitude $340-614$ km). Each constellation uses orbital inclinations and number of satellite planes matching current plans\footnote{\url{https://ls.st/x1o}}. To date, OneWeb has launched and deployed several hundred satellites, while the number of Starlink satellites is in the thousands.

To simulate LSST observations, we start with the baseline observing strategy in \citet{yoachim2022b}. LSST observations are scheduled in visits, where a $u$ visit is one 30s exposure and visits in all other filters ($grizy$) are back-to-back 15s exposures. The baseline strategy attempts to take most observations in mixed filter pairs (e.g., an $r$\ visit followed by an $i$ visit 33 minutes later), and completes 215,000 visits in the first year. 

\begin{figure*}[ht!]
\epsscale{0.37}
\plotone{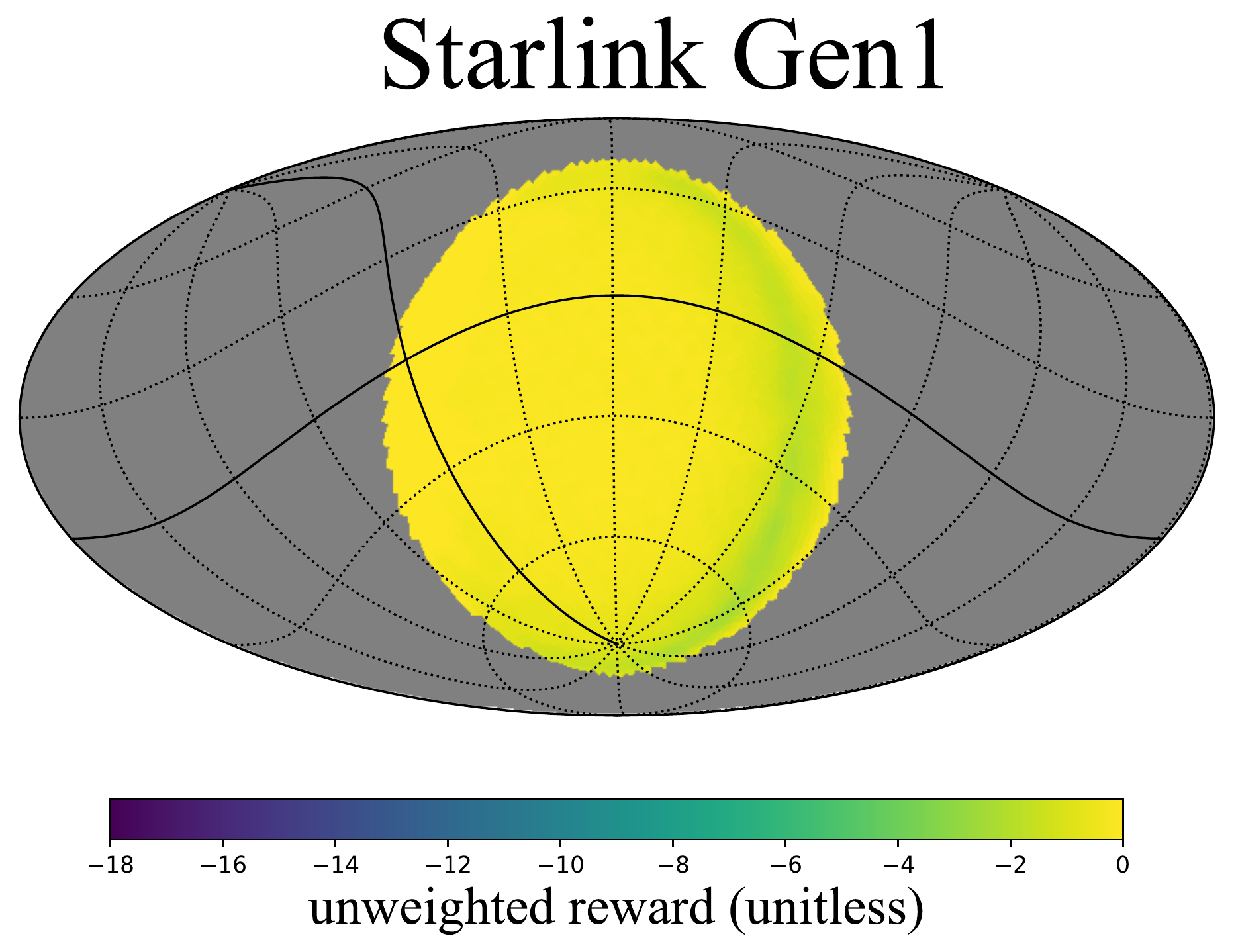}
\plotone{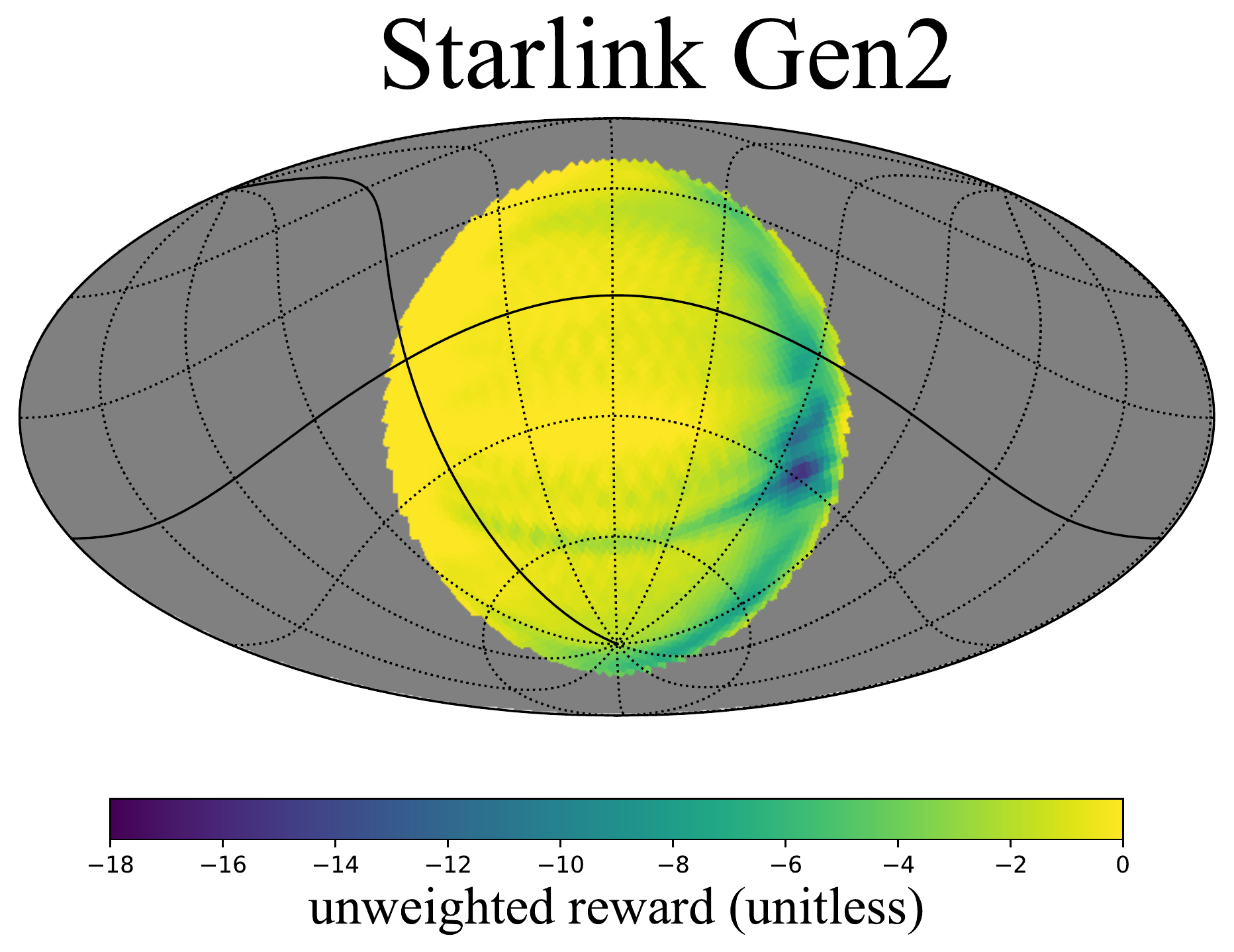}
\plotone{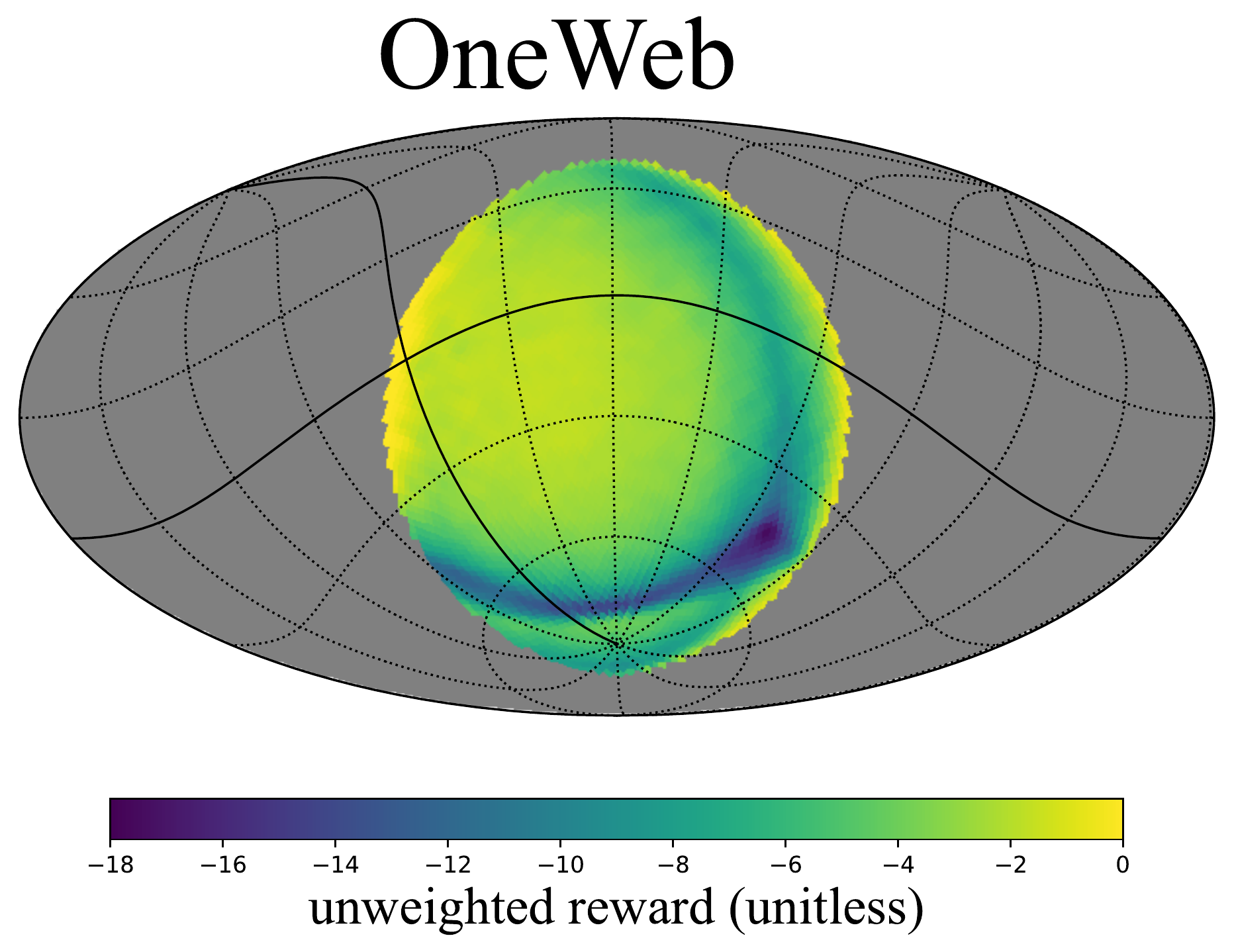}
\epsscale{1}
\caption{Satellite avoidance maps constructed for the Rubin scheduler for each simulated constellation. Each is for a twilight observation period of 90 minutes (beginning after sunset with a Sun altitude of $-17.1$ degrees). The map projections are rotated so zenith is in the center of the image. Darker regions have more illuminated satellites and therefore more negative weighting. By varying the dodging weight placed on these maps, the scheduler will more actively avoid regions of the sky where satellites could streak images.
\label{fig-simulated-scheduler}
}
\end{figure*}

The baseline LSST observing strategy uses three primary basis functions which reward (1) minimizing slewtime, (2) maximizing the depth of images (e.g., by avoiding the Moon and high airmass), and (3) maintaining a uniform survey footprint. To this, we add a fourth basis function which penalizes observing areas of the sky which will have high concentrations of illuminated satellites. Figure~\ref{fig-simulated-scheduler} shows an example of this new basis function for the three simulated satellite constellations. The positions of the illuminated satellites are computed in 10-second intervals and marked on the sky. These maps are then summed over 90-minute blocks to generate the basis function maps. Thus our modified scheduler with a satellite avoidance strategy does not try to avoid individual satellite streaks, but rather has a parameterized method for avoiding regions of the sky where satellite streaks are more likely. This has the additional benefit of not requiring high precision satellite orbit forecasts.

We show three example satellite avoidance maps ready for use by the LSST scheduler in Figure \ref{fig-simulated-scheduler}. It is apparent from Figure \ref{fig-simulated-scheduler} that the simulated OneWeb constellation has more negative area --- regions that should be avoided due to large numbers of illuminated satellites --- than the other two constellations. Although OneWeb has fewer satellites than Starlink Gen2, the OneWeb satellites orbit at a higher altitude (1200 km compared to 340-614 km for Starlink), meaning that they will be illuminated for a longer portion of the night, and also have a larger impact close to twilight. This is why one of the recommendations from \citet{satcon1} is to keep LEO satellites below 600 km altitude.

To investigate whether the scheduler behaves how we expect with the new satellite avoidance strategy, we create a testing function that measures the length of satellite streaks in the simulated fields of view. To ensure efficiency, only satellites that are above the altitude limit and illuminated by the Sun are considered. Satellites below the altitude limit (indicated in the gray region in Figure \ref{fig-simulated-scheduler}) cannot be reached by the Simonyi Survey Telescope and are therefore not included. For each satellite, we first determine whether it is in the field of view for a given pointing by calculating their distance from the center of the field of view. If this distance is less than the radius of the field of view, the satellite has crossed through the pointing. To quantify the impact of the satellite on the pointing, we then project both the satellite location and the pointing to a 2D x,y plane. In this plane, the field of view is roughly circular and the start and end locations of the satellite crossing are two points on the plane, and we can calculate the total intersection length. Therefore, given a simulated satellite constellation and a schedule of observations, we are able to record the number of satellites in each pointing and measure the total streak length. We assume that 
the impact of streaks on science is proportional to their total length, which allows us to quantify the efficiency of the satellite avoidance strategy. 

To estimate the fraction of pixels affected by streaks, following \citet{hasan22} we adopt a fiducial width of 300 pixels (equivalent to 1 arcminute given the plate scale of 0.2 arcseconds per pixel). The 3.5 degree diameter Rubin focal plane is populated with 189 4kx4k CCDs. Assuming a length of 15 CCDs (with a CCD side of 4096 pixels), a single streak corresponds to 0.6\% of all the pixels in the focal plane. On the other hand, if a streak is so bright that entire CCDs are rendered scientifically useless, a single streak would wipe out 8\% of all pixels in the focal plane. 

We simulate observations for only the first year of the planned ten-year LSST, as the survey strategy does not significantly change in later years. We acknowledge this does not account for the likely satellite population increase beyond the three simulated constellations; however, our results should scale linearly to larger future constellations in similar orbital distributions. We do not consider the effects of satellites launching or de-orbiting, and for simplicity we assume each satellite's orbital parameters are constant. This should be an acceptable approximation as long as actual satellite orbital parameters are available $\sim1$\ day in advance so our avoidance basis functions can be constructed. If there is no timely information publicly available on LEO satellite constellation orbits, or it is highly inaccurate, the satellite avoidance strategy would be impossible to implement. We estimate that satellite orbital solutions correct to within about a degree in space and to within a few minutes in time would be sufficient to effectively avoid some regions of the sky with more satellites in large constellations. Observing current Starlink satellites, \citet{Halferty2022} find they can construct TLEs which predict positions with sub-degree spatial precision and sub-second temporal precision which is more than adequate for our proposed satellite avoidance methods.

\section{Results}\label{results}

\begin{figure*}[ht!]
\plottwo{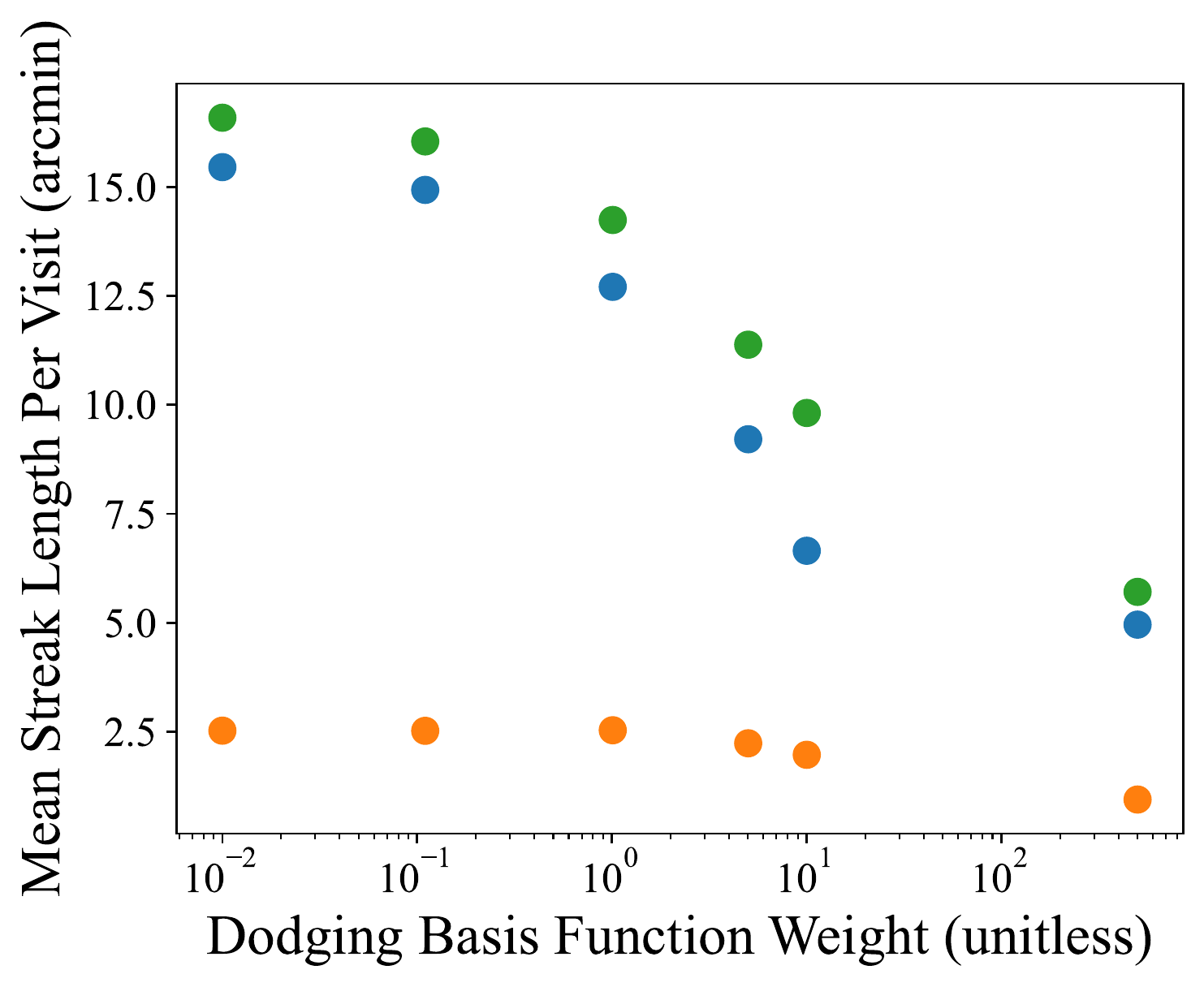}{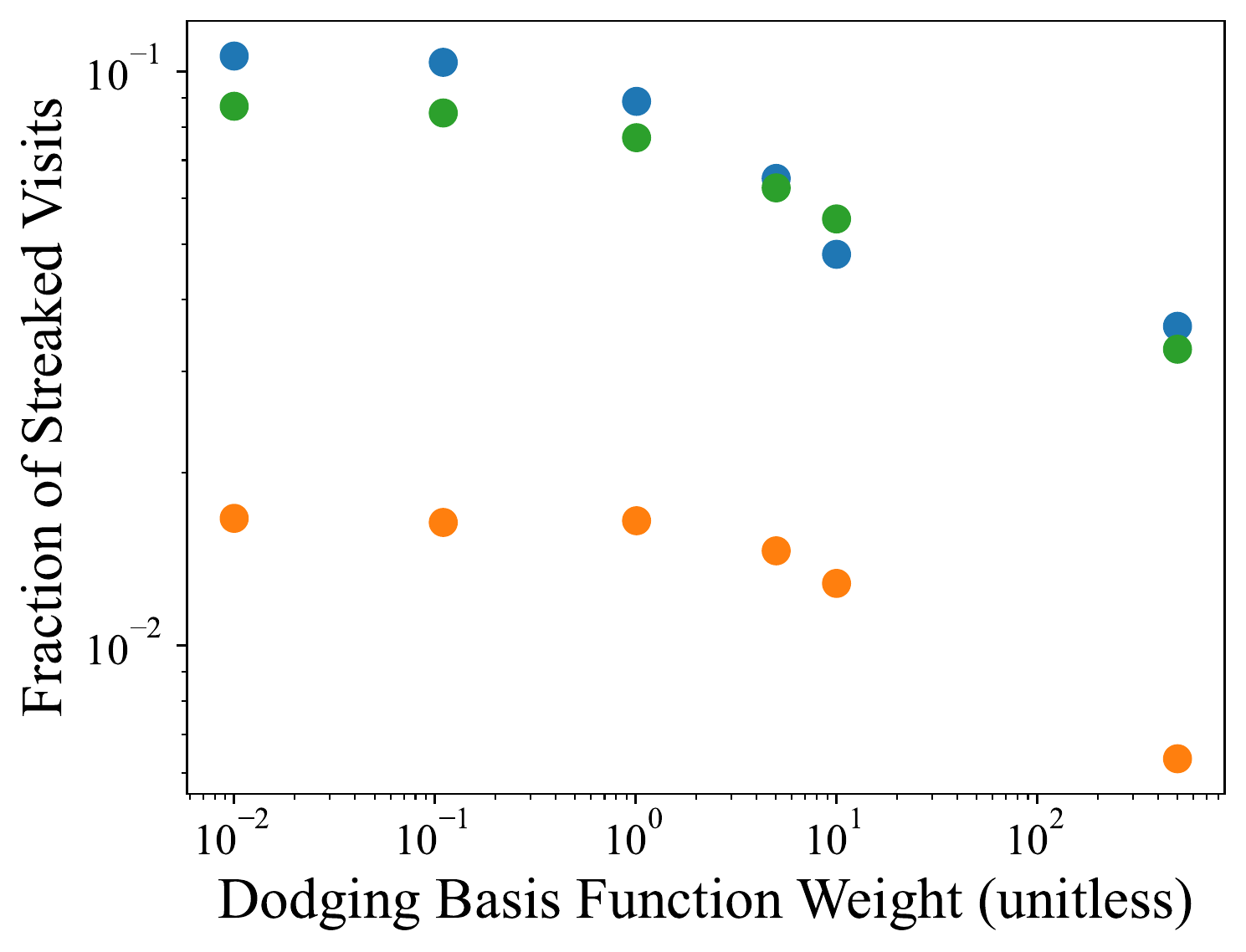}
\plottwo{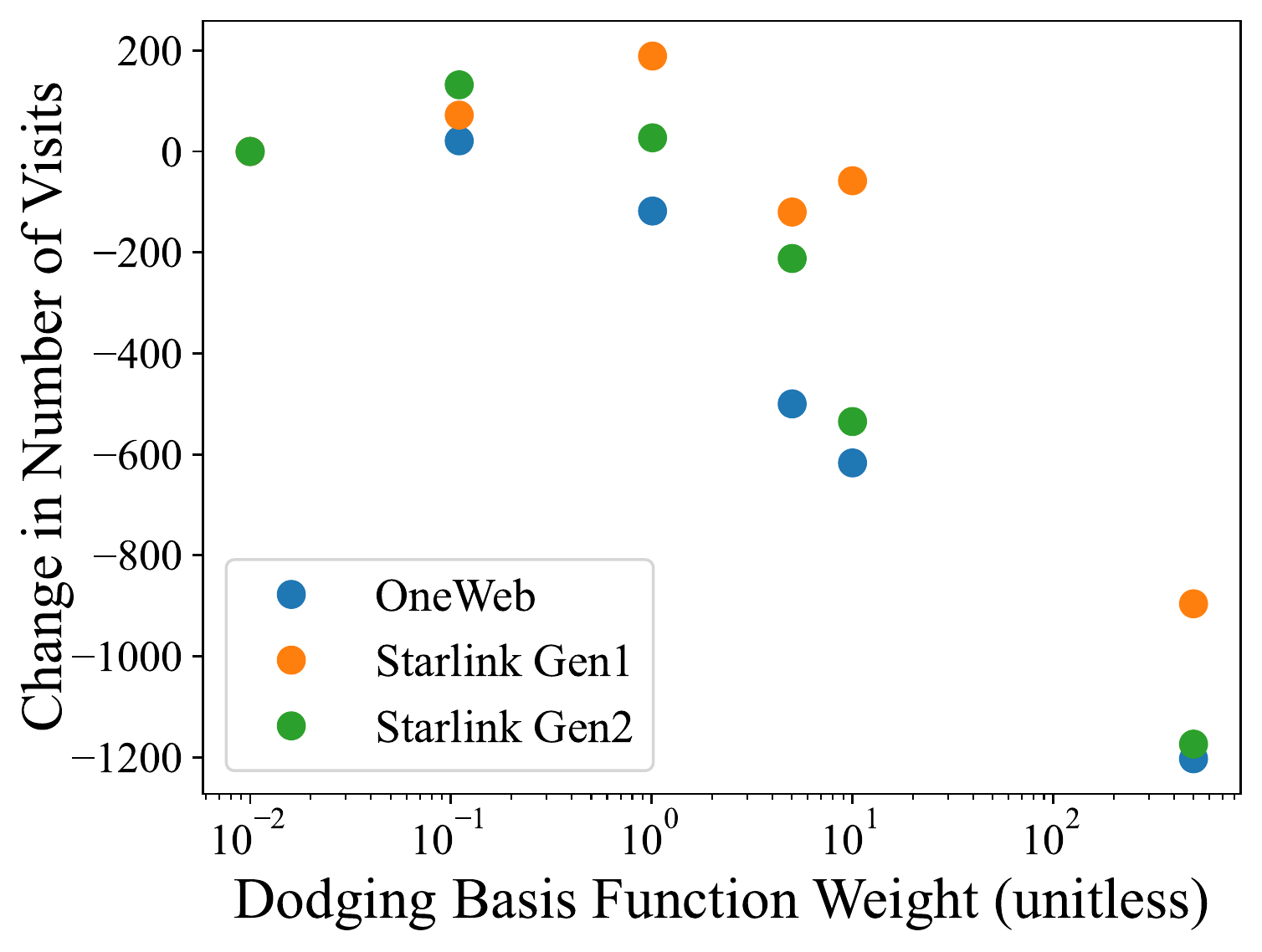}{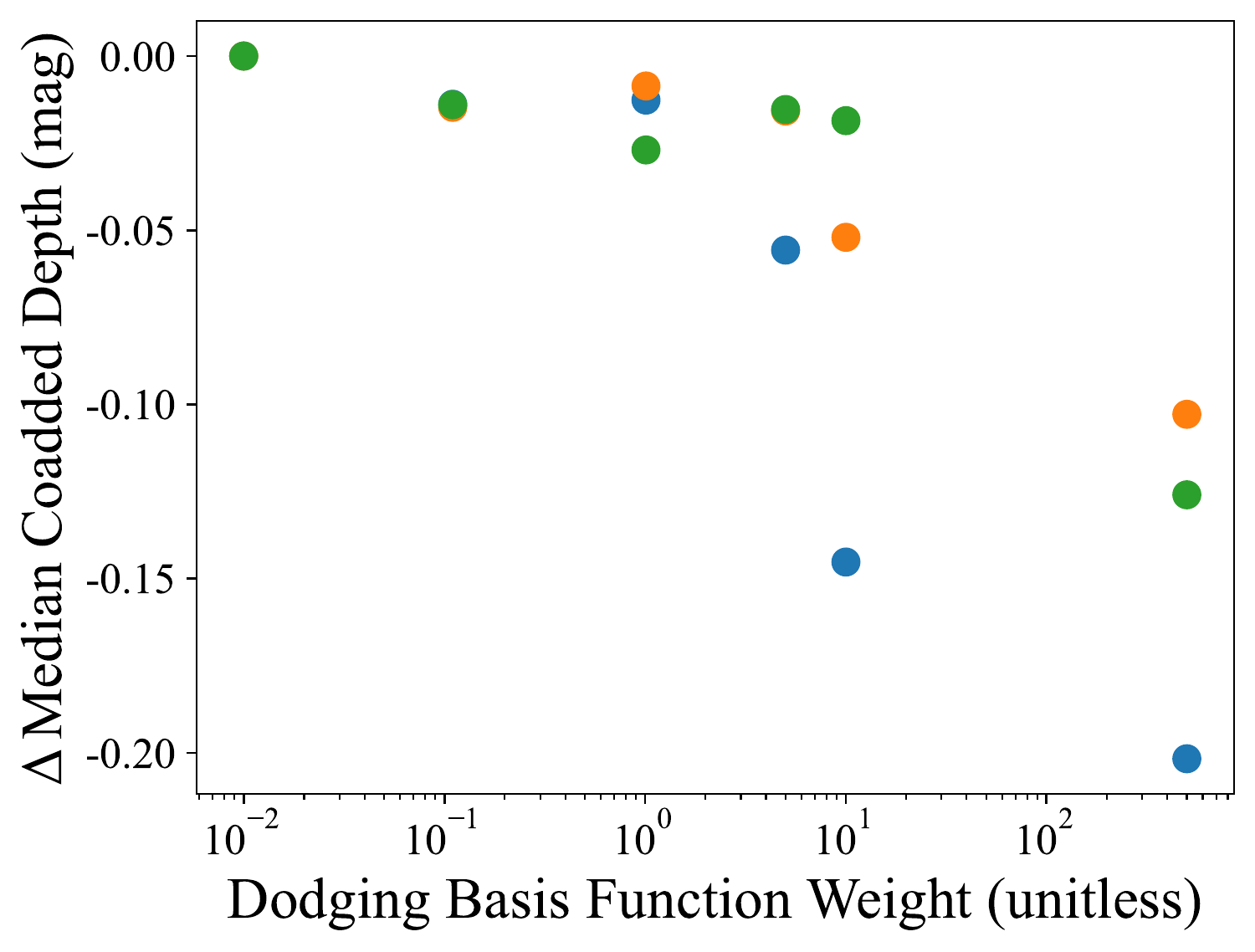}
\epsscale{1.2}
\caption{Illustration of changes in the mean streak length in all visits including those with no streak (top left), the fraction of visits with streaks
(top right), the number of acquired visits in year 1 (bottom left) and coadded depth in the $g$ band (bottom right) as a function of dodging basis function weight (starting with essentially no dodging on the left). 
Note that to reduce the fraction of visits with streaks by about a factor of two, satellite avoidance will require 10\% of total observing time.
\vspace{3em}
\label{fig-pixel-loss-weight}}
\end{figure*}

We find that higher dodging weights reduces pixels lost to satellite streaks, and that the satellite avoidance algorithm is able to effectively avoid satellite streaks in simulated pointings. This is shown in the top two panels of Figure \ref{fig-pixel-loss-weight}. We also find that smaller constellations at lower orbital altitudes (Starlink Gen1, for example) inherently cause less pixel loss per pointing, nearly independent of the dodging weight.

Next, we investigate the relationship between the number of exposures the scheduler is able to complete as a function of the dodging weight. As shown in the bottom two panels of Figure \ref{fig-pixel-loss-weight}, higher dodging weight results in fewer visits, most likely due to longer slew times. With a higher dodging weight, the telescope may be prompted to slew to a location other than the most desirable nearby pointings, resulting in fewer overall exposures. We also find that a larger constellation (Starlink Gen2) tends to decrease the number of exposures slightly more than a smaller constellation (Starlink Gen1), which is expected. More satellites or satellites at higher orbital altitudes result in larger areas of avoidance on the sky, which leads to more slewing required to avoid the affected areas, which subsequently reduces the total number of visits. In addition to forcing longer slews, avoiding satellite dense areas pushes the scheduler to observe pointings with lower signal to noise (e.g., higher airmass, brighter sky background areas) than it normally would.

\begin{figure}[ht!]
\plotone{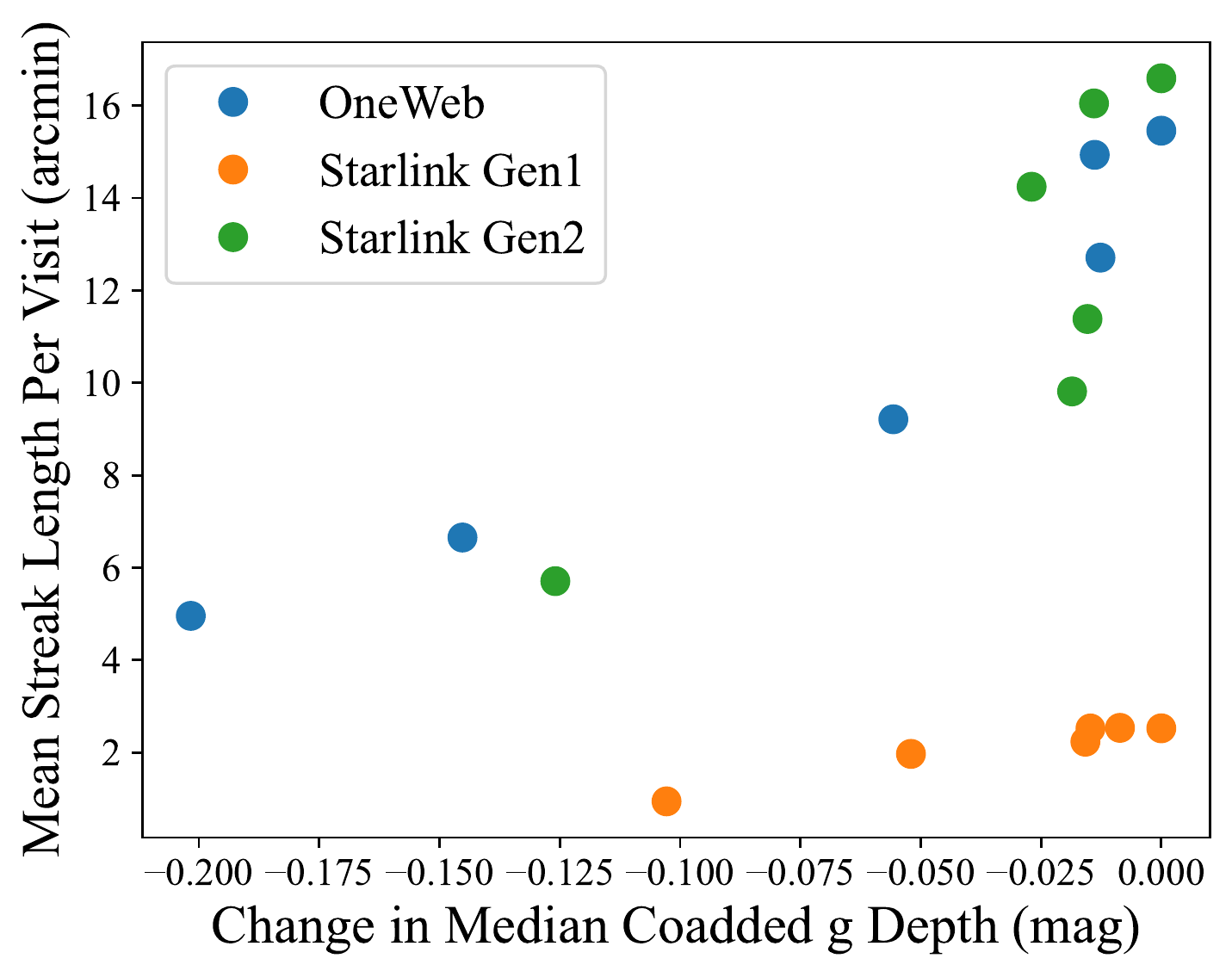}
\epsscale{1.5}
\caption{The trade-off between the mean streak length and final median co-added depth in $g$ band for one year of the LSST (controlled by dodging basis function weight). A negative change in co-added depth indicates the survey is shallower. 
\label{fig-trade-off}}
\end{figure}

One important LSST survey goal is to collect a large number of exposures of the whole southern sky so these may be co-added to reveal faint structures that are not visible in individual visits. As a result, survey depth is crucial to LSST science, and the trade-off between pixel loss from satellite streaks versus survey depth reduction from fewer total visits must be evaluated. With the satellite avoidance algorithm, the scheduler is prompted to avoid regions with illuminated satellites, which sometimes results in longer slew times or less desirable pointing conditions and contributes to a loss in survey depth. Therefore, Figure \ref{fig-trade-off} explores the trade-off between survey depth loss and satellite avoidance. As evident
from the figure, the fraction of LSST visits with streaks can be decreased by a factor of two with 
an investment of 10\% of LSST observing time, corresponding to a loss of coadded depth of 0.05 mag. 

\begin{figure*}
\centering
\epsscale{.85}
\plottwo{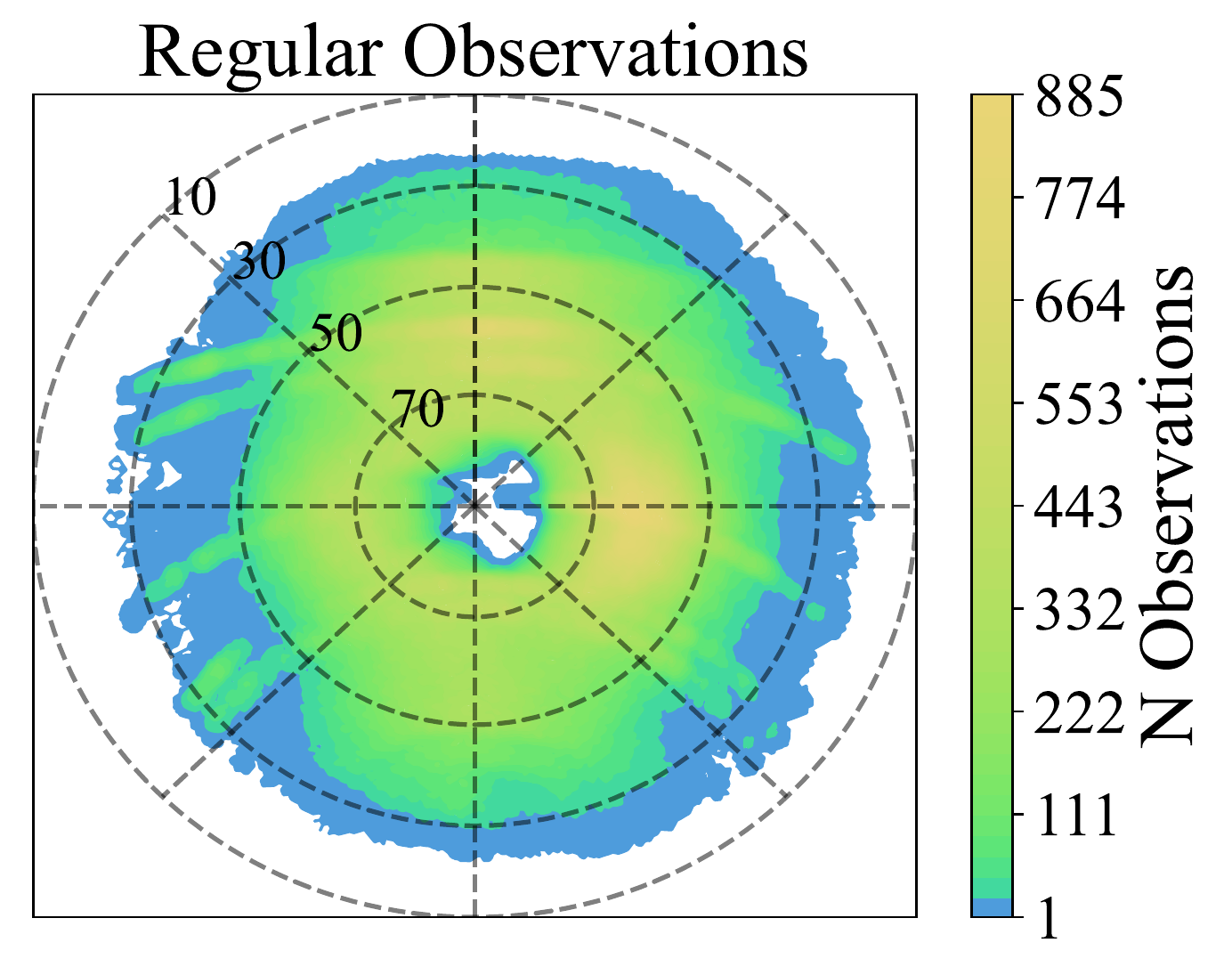}{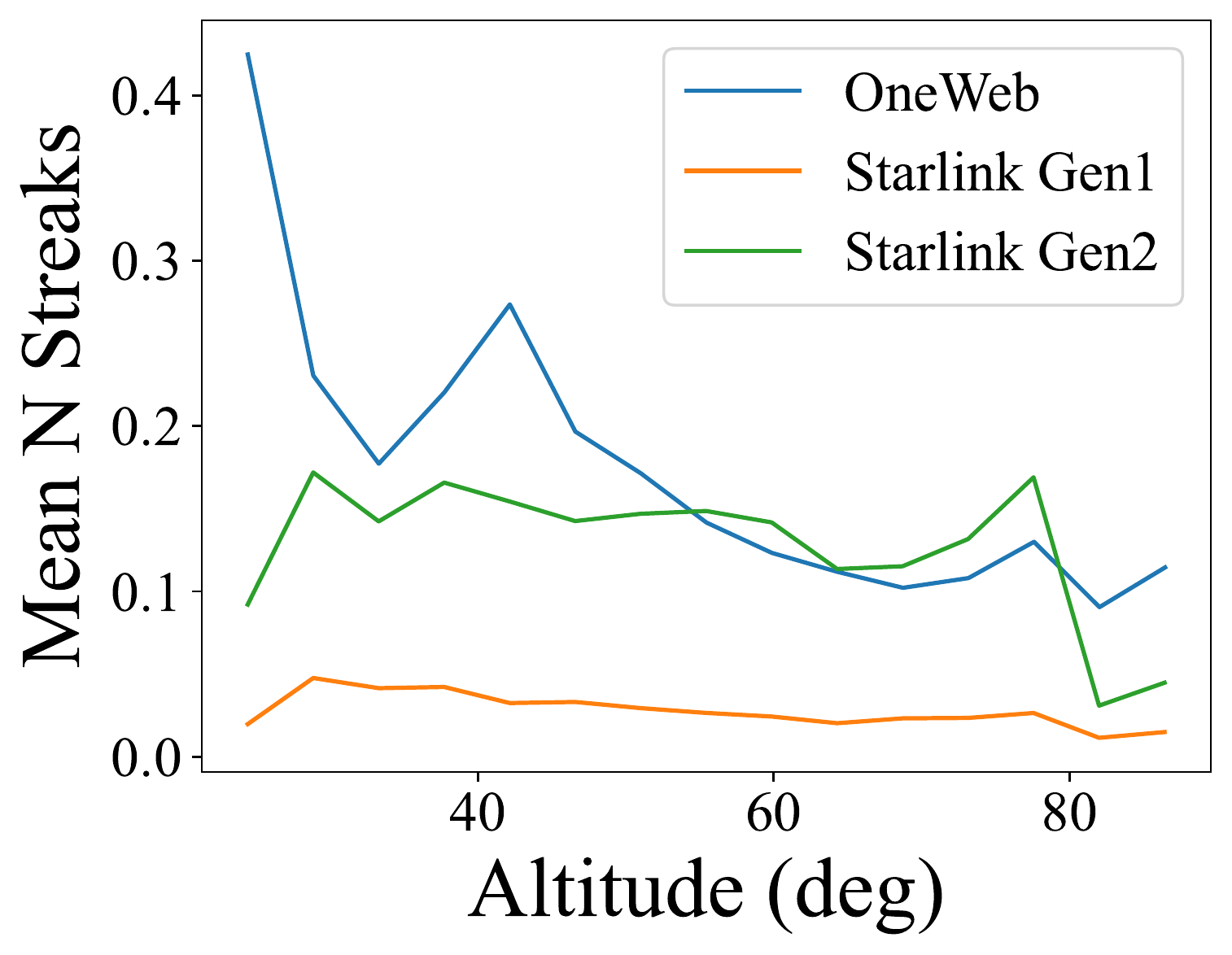}
\plottwo{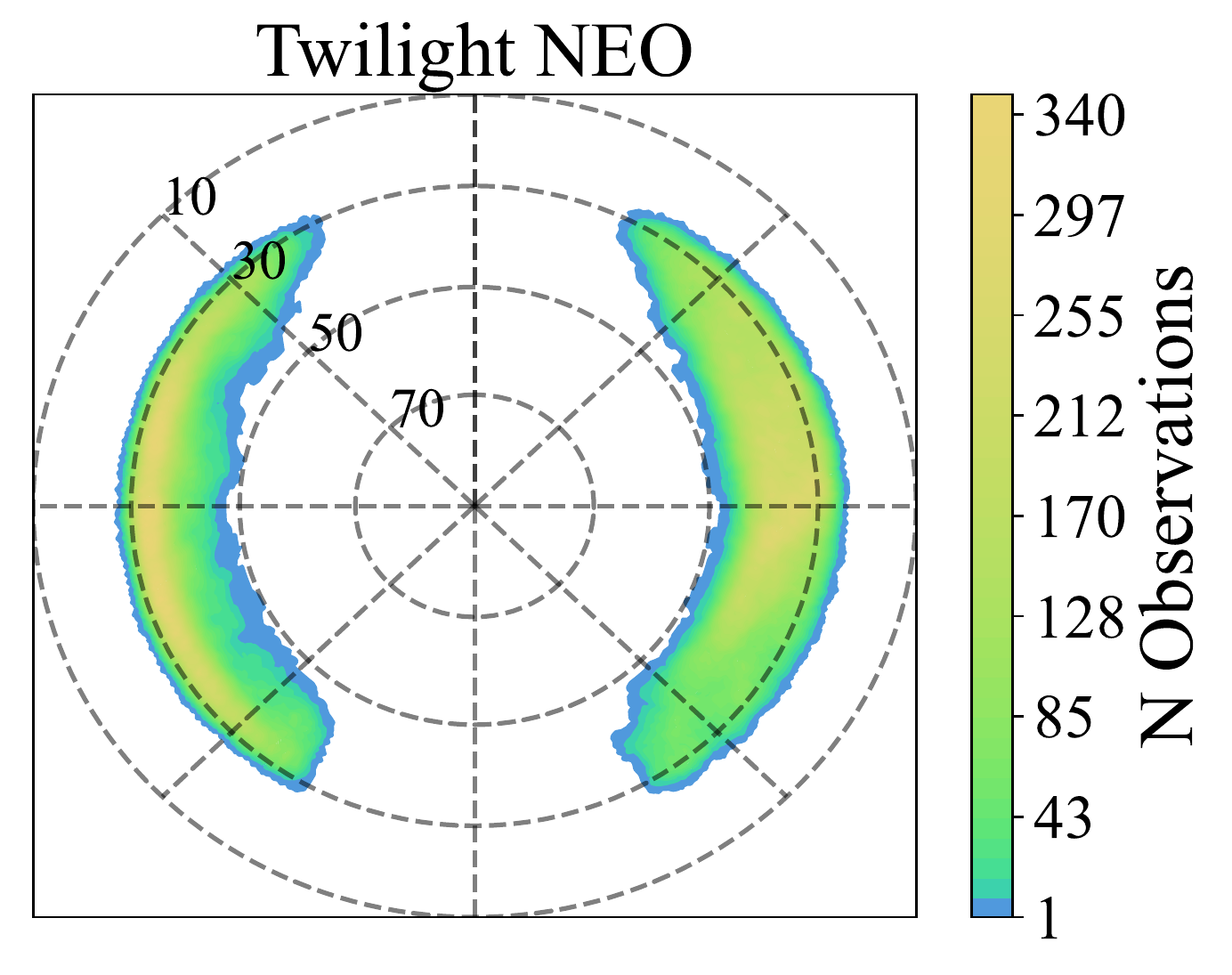}{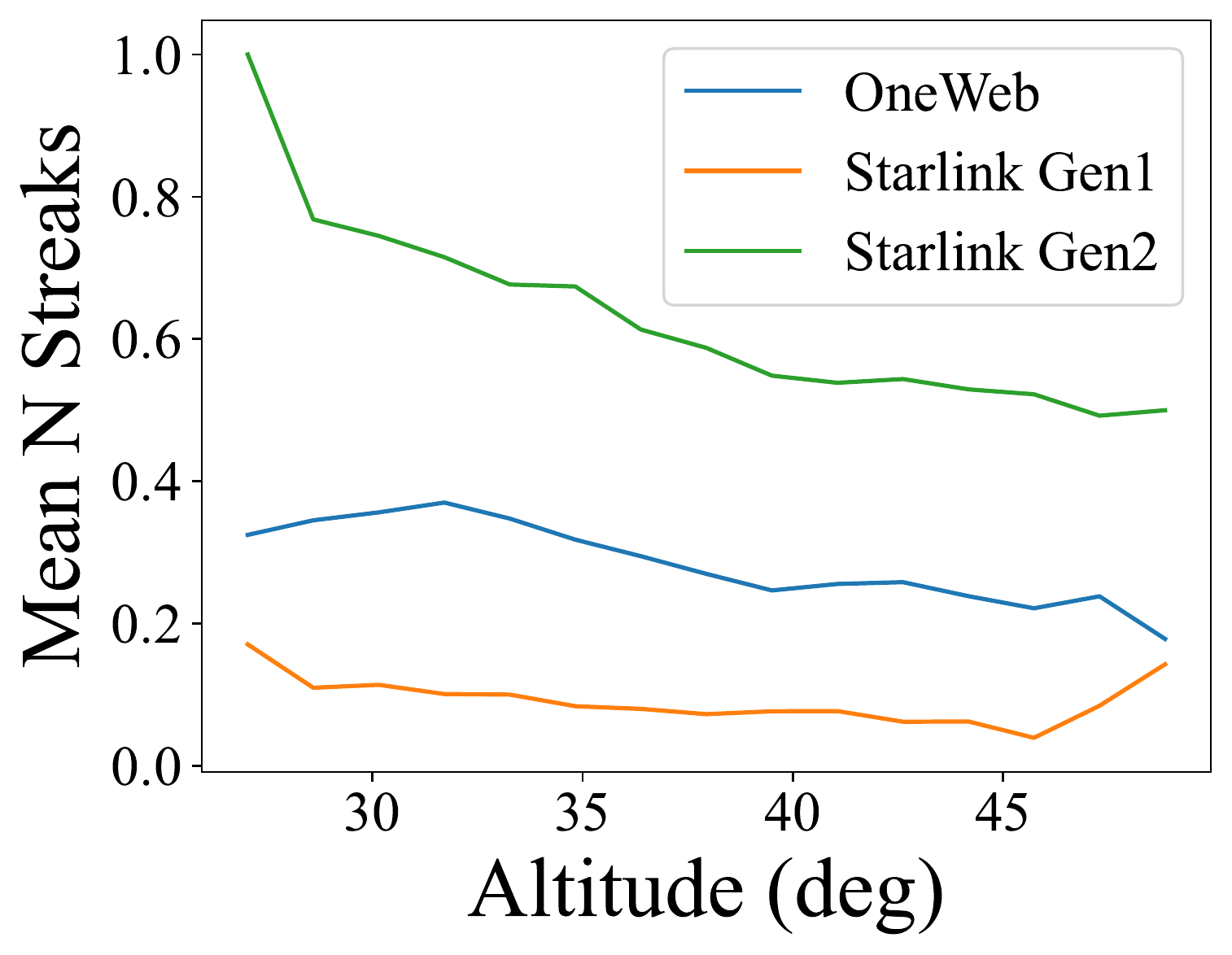}
\epsscale{1}
\caption{Impacts of satellite streaks on simulated LSST observations without any satellite avoidance. Compared to standard LSST observations (top), twilight NEO observations (bottom) cannot easily be shifted to avoid satellites. The left panels show the altitude and azimuth distribution of observations on the sky (zenith at the center of the plots), and the right panels show how many streaks would result from the three simulated satellite constellations as a function of how high above the horizon the telescope is pointing (observation altitude). Note that  most twilight NEO observations would contain a satellite streak.
\label{fig:twi_neo}}
\end{figure*}

So far, we have primarily considered impacts on the overall LSST. However, another important LSST science goal involves using twilight images to search for Near Earth Objects (NEOs) and solar system objects interior to the orbits of Earth and Venus. These observations must be taken in the direction of the rising or setting sun at high airmass. Because a small potential area is targeted, our proposed satellite avoidance scheme is ineffective. Figure~\ref{fig:twi_neo} shows how regular survey observations and twilight NEO observations would be affected by satellite constellations using the LSST scheduler with no satellite avoidance.  While the majority of twilight NEO observations could include a satellite streak with a Starlink Gen2 size constellation, we find this would only result in a $\sim$0.36\% loss of science pixels (for a 1 arcminute wide mask).

\section{Discussion}\label{discuss}

We have demonstrated that adding a weighted term in the scheduler algorithm for illuminated satellites can effectively reduce the amount of satellite streaks in observations, and subsequently reduce mean pixel loss per visit and other impacts on science, as illustrated in Figure \ref{fig-pixel-loss-weight}. 
However, with the new added priority on satellite avoidance, the telescope can be pushed to take an observation path that does not optimize slew time, which subsequently reduces the number of exposures and overall survey depth. The trade-off essentially comes down to the relationship between total streak length reduction and survey depth reduction. The final decision will depend on the overall impact of streaks on science which is not well quantified yet due to a lack of adequate information about the satellite brightness distribution and the impact of glints and low surface brightness residuals on alert purity and systematic errors in cosmological parameter estimation. In other words, 
when evaluating whether to implement a weighted satellite avoidance strategy to effectively reduce satellite streak density, it is necessary to evaluate the trade-off between pixel loss together with non-linear crosstalk, time-domain glint effects, and any other relevant systematics versus loss in observing time. 

We note that earlier publications \citep{lawrence22,tyson20} stated that the majority of LSST images are likely to contain a satellite streak. They also included some higher-altitude Starlink orbits that are no longer planned. Our study finds that about 10\% of all LSST images will have a streak from the three simulated constellations (Starlink Gen1, Starlink Gen2, and OneWeb, totaling 40,768 satellites as currently planned). It is true, however, that twilight observing campaigns at high airmass --- like those necessary to perform NEO searches --- will have streaks in the majority of images.

There is a concerning possibility of sharp increases in satellite population in the next $5–10$ years, overlapping the LSST operations period ($2024–2035$). With a dramatic increase in satellite population, the ability to avoid satellites might become more relevant. 
It is possible to linearly extrapolate our results to consider a possible LEO satellite population in the hundreds of thousands circa 2030, assuming the orbital distribution in LEO is similar to that of Starlink and OneWeb. A future with 400,000 LEO satellites rather than 40,000 --- the stated goal of various companies intending to launch very large constellations given the present filings\footnote{\url{https://planet4589.org/space/stats/conlist.html}} --- could render the trade-off of 10\% of LSST observing time in order to cut the number of visits with streaks in half worthwhile. 
\citet{tyson20} find that satellites with AB magnitudes of $g \sim 3.2$ to $y \sim 1.5$ would saturate LSST images. 
Satellite streaks from Starlink and OneWeb as presently designed are not expected to saturate the LSST Camera's CCD detectors as they have magnitudes of $\sim5$ in their final orbits \citep{Halferty2022}. While a star of $g=10$ would be very saturated in a LSST images, LEO satellites are moving fast enough that their effective exposure time is much lower than astronomical targets. Satellites typically only leave streaks in images when they are both illuminated by the Sun and visible from the observatory, and LEO satellites spend most of the night in Earth's shadow.
Satellites from other operators may be significantly brighter than present-day Starlink and OneWeb satellites, and may saturate the LSST Camera’s detectors or cause overwhelming levels of non-linear crosstalk. In particular, the Blue Walker 3 satellite is the first of a proposed 100 satellite constellation which have a $V$ magnitude between 1-0. Such a bright satellite would saturate LSST images, potentially causing much higher pixel losses than satellites which have been launched to date.
 
Therefore, one future work direction involves adding brightness weighting to the satellite avoidance algorithm. The idea is to only avoid satellites brighter than a certain brightness threshold. This could potentially reduce the region of avoidance, therefore reducing the loss in observing time and co-added depth. It may be possible to compute optimal starting locations for a series of observations based on satellite forecasts to further optimize satellite avoidance. Finally, since faint trail detection and masking is not perfect, no satellite avoidance strategy will effectively mitigate faint glints and the resulting bogus alerts.

\software{Astropy \citep{astropy2013, astropy2018, astropy2022}, 
          Healpy and HEALpix\footnote{\url{http://healpix.sourceforge.net}} \citep{healpix2005, Zonca2019},
          Matplotlib \citep{Hunter:2007},
          Numpy \citep{harris2020array},
          rubin\_sim \citep{yoachim2022},
          Scipy \citep{2020SciPy-NMeth},
          Skyfield \citep{Rhodes2019}, 
          Shapely \citep{shapely2007}}

\begin{acknowledgments}
We thank the anonymous referee who provided helpful prompt feedback that improved the clarity of the paper.
JH acknowledges support from the Computing Research Association-Widening Participation (CRA-WP) Distributed Research Experiences for Undergraduates (DREU) program.
JH and MLR are grateful for LSST Corporation travel support for JH to attend the 2022 Rubin Project and Community Workshop and present a poster.
MLR acknowledges support from Bob Blum, Leanne Guy, and all of Rubin Operations to spend a fraction of her time on satellite mitigation work; this study would not have been possible without formal recognition that the proliferation of bright commercial LEO satellites poses a threat to LSST science.
The authors all wish to thank Tony Tyson for valuable discussions and feedback that helped place this work in context.
This work was facilitated through the use of advanced computational, storage, and networking infrastructure provided by the Hyak supercomputer system at the University of Washington.
\end{acknowledgments}

\bibliographystyle{aasjournal}

\end{document}